# Does the Universe Have a Handedness?

Michael J. Longo

University of Michigan, Ann Arbor, MI 48109-1120

In this article I study the distribution of spiral galaxies in the Sloan Digital Sky Survey (SDSS) to investigate whether the universe has an overall handedness. A preference for spiral galaxies in one sector of the sky to be left-handed or right-handed spirals would indicate a preferred handedness. The SDSS data show a strong signal for such an asymmetry with a probability of occurring by chance ~3.0 x $10^{-4}$. The asymmetry axis is at $(RA,\delta)$ ~(202°,25°) with an uncertainty ~15°. The axis appears to be correlated with that of the quadrupole and octopole moments in the WMAP microwave sky survey, an unlikely alignment that has been dubbed "the axis of evil". Our Galaxy is aligned with its spin axis within 8.4° of this spiral axis.

Subject headings: cosmological parameters---galaxies: general---galaxies: spiral---large-scale structure of the universe

## 1. INTRODUCTION

Symmetry has a strong appeal to the human psyche. Nature, however, exhibits some surprising asymmetries. On the smallest scales, an asymmetry (parity violation) was found in the angular distribution of electrons in the beta decay of spin oriented $^{60}$Co, confirming the proposal by Lee and Yang (1956) that parity was violated in weak decays. On the molecular scale, there is a large predominance of left-handed amino acids over right-handed ones in organisms, the origin of which is still not well understood. It is reasonable to ask if nature exhibits such an asymmetry on the largest scales.

The SDSS DR5 database (Adelman-McCarthy et al. 2007) contains ~40000 galaxies with spectra for redshifts <0.04 (corresponding to a distance of about 172 Mpc) with a wide coverage in right ascensions ($RA$) and a more limited range of declinations ($\delta$). A few percent of the gal-axies are spiral galaxies with a clearly recognizable handedness that could be used for this study. A universe with a well-defined handedness is illustrated schematically in Fig. 1(a). Note that to us, galaxies in one hemisphere would be right-handed while they would be left-handed in the opposite hemisphere. (If a predominance of left- or right-handedness were seen in all directions it would be an indication of a bias or systematic error preferring that handedness.) Note that only the component of the spin along our line of sight can be observed. In this ideal universe, our own galaxy would have the same handedness as the other spiral galaxies. Figures 1(b,c) show typical spiral galaxies from the SDSS and gives our convention for "spin" directions.

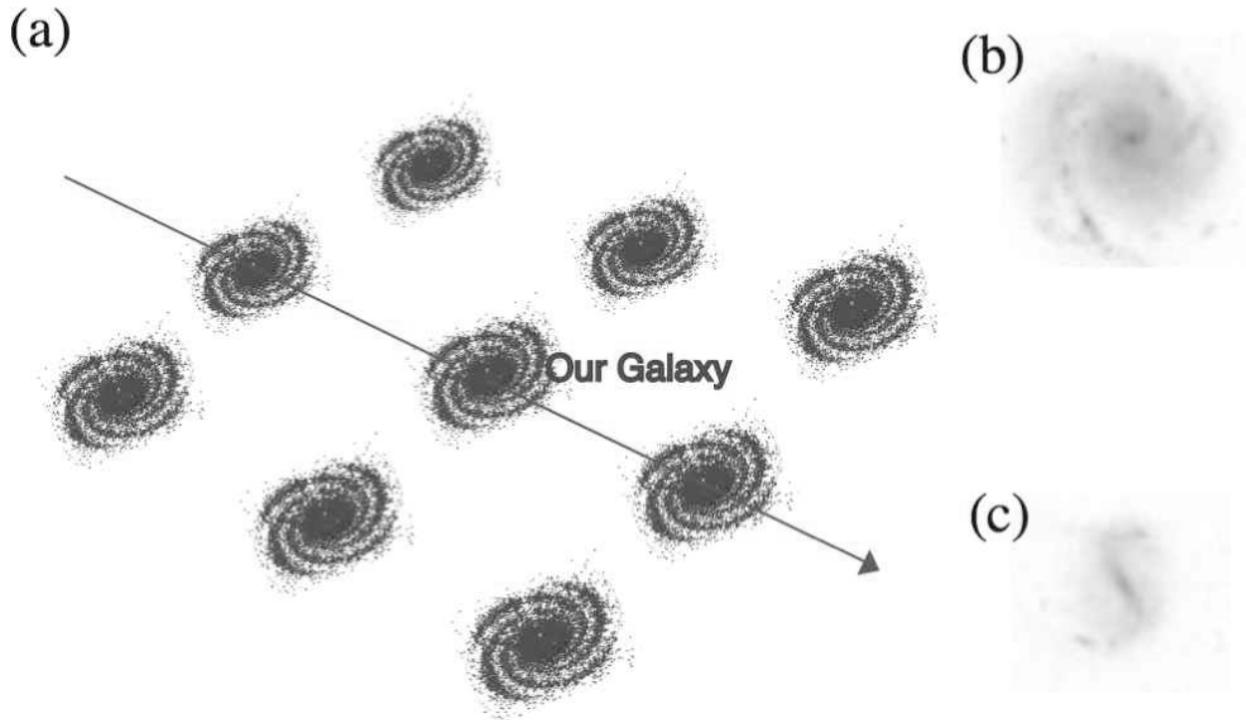

Fig. 1. (a) Perspective view of a hypothetical universe with all galaxies having the same handedness. Note that galaxies in one hemisphere would appear to us to be right-handed and in the opposite hemisphere left-handed. (b) A "typical" spiral galaxy from the SDSS. This one is defined as having right-handed "spin". (c) A left-handed two-armed spiral galaxy.

Considerable effort has gone into looking for alignment of galaxies with local tidal shears. [See, for example, the recent article of Lee and Erdogdu and articles cited there.] These use small samples (only 169 spirals in the case of Lee & Erdogdu).  However, tidal torques can only cause an alignment of galaxy spins on the scale of galaxy clusters.   In an N-body simulation, Porciani et al. (2002) show that the correlations, which are ~5%, disappear beyond a scale of 1 – 8 Mpc. On the scale of 170 Mpc studied here, the effects of tidal torques will be diluted by a factor $\sim(8/170)^3 \sim 10^{-4}$, while the observed spin asymmetry, as discussed below, is $\approx 8\%$. To the author's knowledge, this is the first search for a predominant spin handedness over the whole sky out to relatively large redshifts, rather than a local spin alignment such as that produced by tidal shear on the scale of galaxy clusters.  Also, unlike many other properties of galaxies, their handedness is unaffected by incompleteness of surveys or atmospheric effects.

Figure 2(a) shows the distribution of redshifts as a function of  *RA* for SDSS galaxies with spectra.  As can be seen, the coverage in *RA* is substantial, except for the region obscured by the Galactic disk.  The well-known "foam-like" structure is apparent, even in this thick slice. The distribution of redshifts as a function of declinations is shown in Fig. 2(b).  Here the coverage is spotty, especially in the hemisphere toward $RA = 0°$.

## 2.  THE ANALYSIS

The handedness analysis required galaxies with spectra so that their redshifts could be determined. This  analysis was limited to spiral galaxies with redshifts <0.04 because beyond that it becomes progressively harder to distinguish the handedness of the spirals.  The green apparent magnitude was required to be <17 for the same reason.  A list of galaxies that satisfied these criteria was obtained from the SDSS DR5 web site, cas.sdss.org/dr5.  This yielded 22,704 spiral candidates.  Galaxies from this list were scanned manually to select those that were reasonably clear spirals.  The scanning was done in random order with respect to *RA* and $\delta$ so that any scanning bias could not cause a systematic bias on the (*RA*, $\delta$) distribution of the handedness. This preliminary scan yielded 2817 good spirals that were then downloaded as JPEG files to be analyzed. The *RA*, $\delta$, and *z* coordinates, as well as the magnitudes in 5 spectral bands, were also obtained.

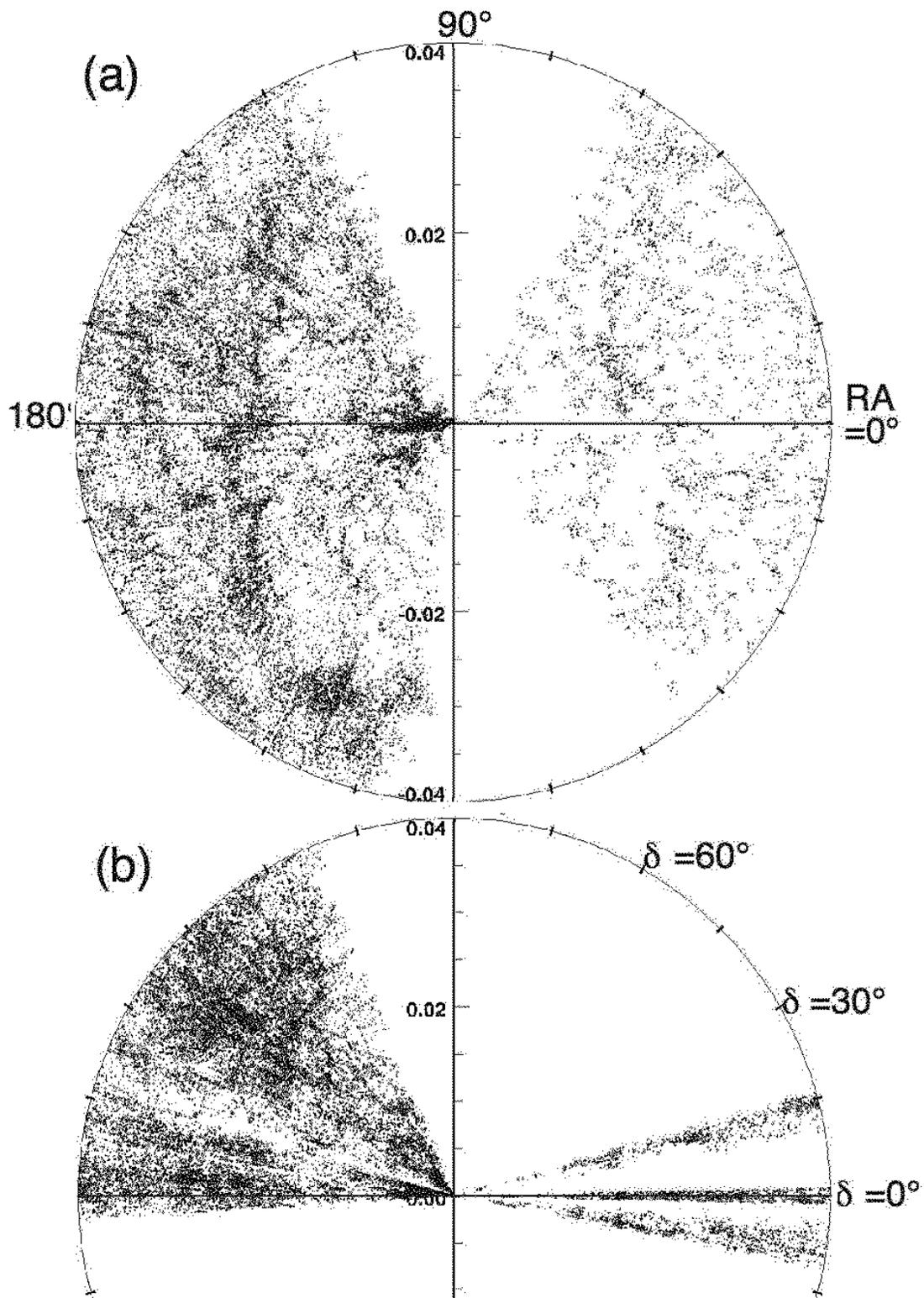

Fig. 2. (a) Polar plot of right ascensions vs. redshifts for all declinations. (b) Plot of declinations vs. redshifts. The left and right hemispheres in (a) are plotted on their respective sides in (b). Declinations between -19° and +60° were used in most of this analysis.

In an earlier version of this analysis (Longo, 2007) the JPEG file for each spiral galaxy was submitted to an IDL program to determine its handedness. The algorithm used for this analysis was based on a rotating one-armed spiral mask similar in shape to half of the spiral in Fig. 1(c). However, the algorithm was developed using spirals at fairly low redshifts, and it was belatedly realized that its efficiency at larger redshifts was poor. Visual scanning was found to be considerably more efficient and it was easy to take precautions to eliminate any human bias toward left- or right-handedness. The results of the two analyses are in complete agreement. Only the visual scan results that include $\approx 50\%$ more galaxies will be presented here.

In the visual scan the spirals were mirrored randomly to prevent scanning bias. Each JPEG file was viewed and labeled as either *L*eft, *R*ight, or *U*nknown. This gave $R = 1256$, $L = 1360$, and $U = 201$.

Many spiral galaxies have undergone collisions with other galaxies since their formation. Such collisions tend to mix "orbital" angular momentum of the galaxies with any primordial spin angular momentum they may have formed with. Collisions between galaxies trigger prolific star formation. Galaxies with recent star formation tend to be bluish. In order to enhance any signal for a preferred handedness, the bluest galaxies were removed from the sample. This selection was imposed by requiring that the ultraviolet and far infrared magnitudes satisfy the condition (*U-V*) > 1.6. It removed 3.2% of the *R+L* spirals.

## 3. RESULTS

A plot of asymmetries $A \equiv (R-L)/(R+L)$ in sectors of right ascension and segments in $z$ is shown in Fig. 3. Positive $A$ are shown in red and negative ones in blue. The larger numbers near the perimeter give the net asymmetry for the entire $RA$ sector. The black numbers in parentheses below them give the total number of galaxies in that sector.

The three sectors between $RA=150°$ and $240°$ appear to show a clear preference for negative asymmetries. The populations in the opposite $RA$ sectors between $+80°$ and $-80°$ are much sparser, but they do show an excess of positive asymmetries. Table I gives for those two $RA$ ranges the number counts, $N^+$ and $N^-$, the asymmetry $\langle A \rangle$, its standard deviation, $\sigma(\langle A \rangle)$ and the ratio $\langle A \rangle / \sigma(\langle A \rangle)$. The $\sigma$ are determined from standard normal distribution statistics, $\sigma(N) = \sqrt{N}$, which gives $\sigma(\langle A \rangle) = 1/\sqrt{N^+ + N^-}$. The asymmetry for $150° < RA < 240°$ differs from zero by 3.11 $\sigma$. The probability for exceeding $3.11\sigma$ by chance is 0.19%. For $-80° < RA < 80°$, the asymmetry is positive, consistent with a signal for a preferred handedness.

In order to study the dependence of the asymmetry on $RA$ and $\delta$ we use only the galaxies between 90° and 270° because of the very limited declination coverage in the hemisphere toward $RA=0°$ (Fig. 2b). The data were first binned in 5 slices in $\delta$ using this entire $RA$ range. Table II shows the asymmetry for the 5 slices in $\delta$. The spin asymmetries in the 5 slices are statistically independent so that the overall probability is the product of that for the 5 slices, $3.0 \times 10^{-4}$. A least-squares fit of the asymmetries to $A_0 \cos(\delta - A_1) + A_2$ gave $A_1 = 25 \pm 7°$ as the declination at which the asymmetry is maximal. A similar fit to slices in *RA* with $5° < \delta < 45°$ gives the *RA* that maximizes the asymmetry as $202 \pm 16°$. However, these uncertainties in $\delta$ and *RA* may be underestimated, as they do not take into account the appropriateness of the fitting functions.

Table I. Number counts and net asymmetries for the $RA$ ranges indicated. The last two columns give the number of standard deviations for the asymmetries and the probability of exceeding that $\langle A \rangle / \sigma$.

| $RA$ Range | $N^+$ | $N^-$ | $N_{Tot}$ | $(N^+ - N^-)/N_{Tot} \pm \sigma$ | $\langle A \rangle / \sigma$ | Probability |
|---|---|---|---|---|---|---|
| 80° to −80° | 159 | 153 | 312 | 0.0192±0.0566 | +0.34 | 0.734 |
| 150° to 240° | 650 | 767 | 1417 | −0.0826±0.0266 | −3.11 | 0.0019 |

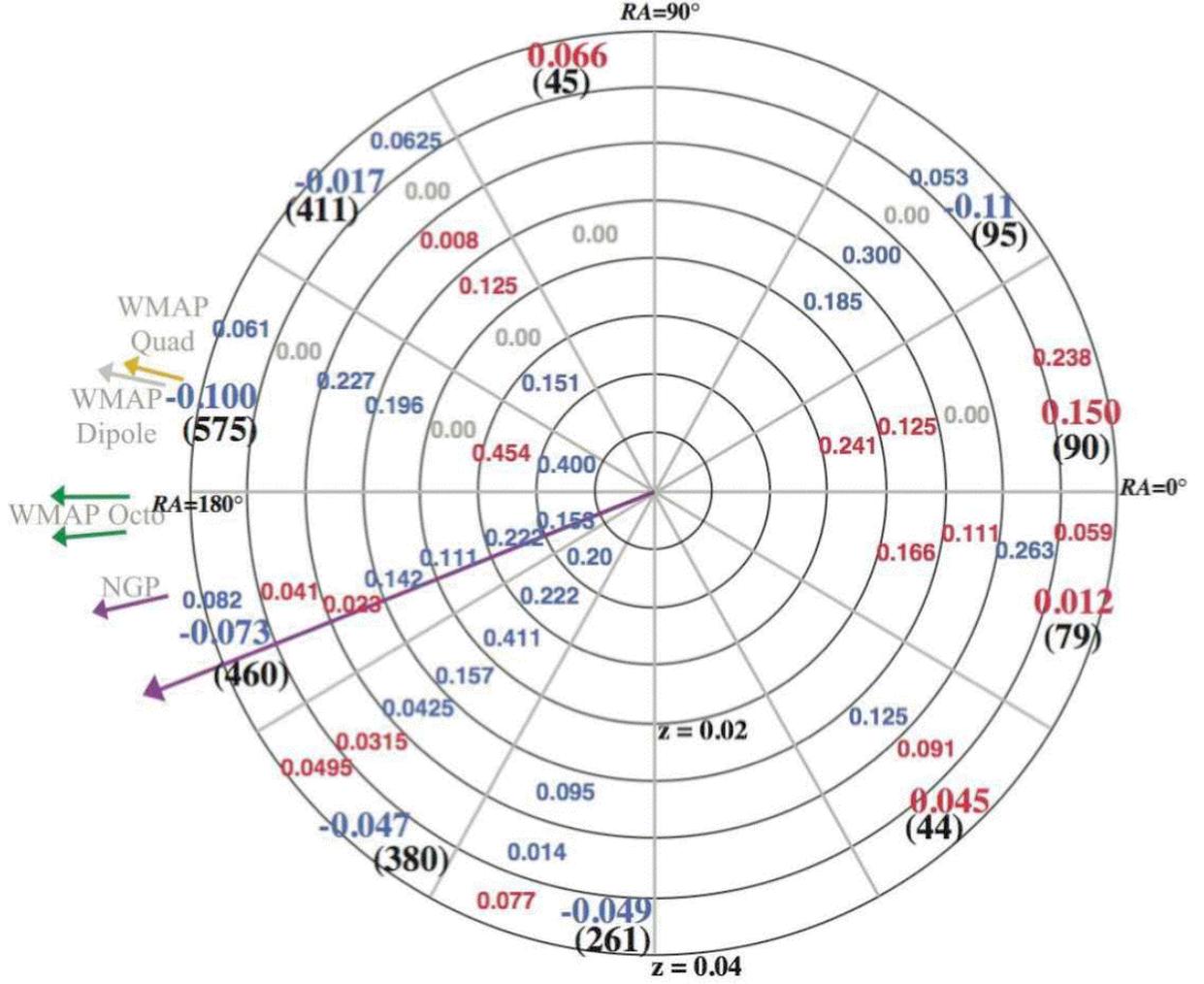

**Fig. 3.** Net asymmetries <A> by sector in *RA* and segments in *z*. Segments with positive <A> are indicated in red and negative <A> in blue. The <A> for segments with <10 galaxies are not shown. The larger numbers near the periphery give the overall asymmetry for that sector; the black numbers in parentheses are the total number of spiral galaxies in the sector. The arrows on the left show the direction of the WMAP quadrupole and octopole moments. The NGP is the north pole of our Galaxy. The long purple arrow shows the most probable spiral spin axis. Declinations between -19° and +60° were used.

Table II. Net asymmetries for the $\delta$ ranges indicated. The last two columns give the number of standard deviations for the asymmetries and the probability of exceeding that $\langle A \rangle / \sigma$.

| $\delta$ Range | $(N^+ - N^-)/N_{Tot} \pm \sigma$ | $\langle A \rangle / \sigma$ | Probability |
|---|---|---|---|
| −10° to +4° | −0.0584±0.0570 | −1.03 | 0.305 |
| +4° to 18° | −0.0104±0.0510 | −0.20 | 0.838 |
| +18° to 35° | −0.1062±0.0497 | −2.14 | 0.0326 |
| +35° to 45° | −0.0906±0.0434 | −2.09 | 0.0371 |
| +45° to 65° | −0.0017±0.0416 | −0.04 | 0.966 |

## 4. DISCUSSION

The Wilkinson Microwave Anisotropy Probe (WMAP) studied the cosmic microwave background (CMB) radiation (G. Hinshaw et al. 2006). Their results for the angular power spectra have been analyzed by Schwarz et al. (2004) and many others. Schwarz et al. show that: *(1)* the quadrupole plane and the three octopole planes are aligned, *(2)* three of these are orthogonal to the ecliptic, *(3)* the normals to these planes are aligned with the direction of the cosmological dipole and with the equinoxes. The respective probabilities that these alignments could happen by chance are 0.1%, 0.9%, and 0.4%. This alignment is considered to be so bizarre that it has been referred to as "the axis of evil" (AE) by K. Land and J. Magueijo (2005). Their nominal AE is at $(l, b) \approx (-100°, 60°)$, corresponding to $(RA, \delta) = (173°, 4°)$. The alignment with the ecliptic and equinoxes is especially problematic because this would suggest a serious bias in the WMAP data that is related to the direction of the Earth's spin axis, which is highly unlikely. Resolving this quandary requires data from another source with different systematics than WMAP.

On the basis of the 3-year WMAP data (ILC123), Copi et al. (2007) find the normal to the plane defined by the quadrupole moments to be at $l = -128.3°$, $b = 63.0°$ in Galactic polar coordinates. In equatorial coordinates this corresponds to $RA = 166°$, $\delta = 16°$ compared to $RA \approx 202°$, $\delta \approx 25°$ for the spin asymmetry axis seen here. (See Fig. 3.) The spin asymmetry axis is also close to two of the three octopole axes at $(RA, \delta) = (181°, -11°)$ and $(185°, 38°)$.

We also see that the spin asymmetry is well aligned with the North Galactic Pole of our Galaxy (NGP in Fig. 3) at (193°,27°). This is due to the fact that our Galaxy also has its axis along the spin alignment with a handedness like that of the majority of the spiral galaxies. These two axes are 8.4° apart. The probability that this came about by chance is 0.5%. Since most astronomical surveys, including the SDSS, tend to make observations toward the NGP and SGP to avoid the obscuration due to the Milky Way, the spin alignment axis is *accidentally* aligned with the SDSS coverage.

The approximate agreement of the spin alignment axis with the WMAP quadrupole/ octopole axes reinforces the finding of an asymmetry in spiral galaxy handedness and suggests that this special axis spans the universe. The fact that the spin asymmetry appears to be independent of redshift suggests that it is not connected to local structure. On the other hand, the spiral galaxy handedness represents a unique and completely independent confirmation that the AE is not an artifact in the WMAP data due to foreground contamination.

Campanelli, Cea, and Tedesco (2006) have suggested that an ellipsoidal universe, possibly the result of a uniform cosmic magnetic field $\sim 10^{-9}$ G, could explain the anomalies observed in the low-$\ell$ amplitudes of the WMAP data. The mechanism that produces the alignment of the low-$\ell$ multipoles could yield a dipole that overwhelms the dipole moment induced by our motion through the CMB rest frame, causing the apparent alignment of the dipole axis with the AE (Fig. 3).

The question of the origin of astrophysical magnetic fields remains controversial. Kulsrud and Zweibel (2007) give a recent review with over 100 references. The observational evidence suggests that the magnetic field is already moderately strong at the beginning of the life of a galaxy. Kulsrud and Zweibel conclude that it is likely the fields are first produced at moderate strengths by primordial mechanisms that have not yet been seriously developed. Since interga-lactic matter is mostly ionized, a possible scenario for producing a preferred handedness is that in the earliest stages of galaxy formation the primordial magnetic field causes electrons and positive ions to spiral in opposite directions around the magnetic field lines. The electrons can shed angular momentum by radiation, so that the final angular momentum of the galaxy is determined by that of the protons.

Perhaps more importantly, a well-defined axis for the universe on a scale ~170 Mpc would mean a small, but significant, violation of the Cosmological Principle and of Lorentz symmetry and thus of the underpinnings of special and general relativity. It is interesting to note that the preferred handedness of spiral galaxy implies that the universe has a handedness as well as a unique axis so that *CP* and parity are violated on the largest scales.

I am extremely grateful to the SDSS group whose efforts and dedication made this work possible.

# REFERENCES

Adelman-McCarthy, J., et al., 2006, ApJS, 162, 38

Campanelli, L., Cea, P., & Tedesco, L. 2006, Phys. Rev. Lett., 97, 131302

Copi, C., Huterer, D., Schwarz, D.J., & Starkman, G.D. 2007, Phys. Rev. D, 75, 023507

Hinshaw, G., et al. 2007, ApJS, 170, 288

Kulsrud, R. & Zweibel, E, 2007, preprint (astro-ph/0707.2783), invited review for Phys. Rep.

Land, K. & Magueijo, J. 2005, Phys. Rev. Lett., 95, 071301

Lee, J. & Erdogdu, P., 2007, preprint (astro-ph/0706.1412), submitted to ApJL

Lee, T.D. & Yang, C.N. 1956, Phys. Rev., 104, 254

Longo, M. J., 2007, preprint (astro-ph/0703325)

Porciani, C., Dekel, A., & Hoffman, Y., MNRAS, 332, 325

Schwarz, D.J., Starkman, G.D., Huterer, D., & Copi, C.J. 2004, Phys. Rev. Lett., 93, 221301